# Gamma-Ray Pulsar Discovery Using a Lattice-Based Method

Daniel Pook-Kolb,[1,2] Dotan Gazith,[3] Colin J. Clark,[1,2] Erik Agrell,[4] Barak Zackay,[3] and Bruce Allen[1,2]
[1]*Max Planck Institute for Gravitational Physics (Albert Einstein Institute), Callinstraße 38, 30167 Hannover, Germany*
[2]*Leibniz Universität Hannover, 30167 Hannover, Germany*
[3]*Department of Particle Physics and Astrophysics, Weizmann Institute of Science, 76100 Rehovot, Israel*
[4]*Department of Electrical Engineering, Chalmers University of Technology, 41296 Gothenburg, Sweden*

Pulsars are difficult to discover via $\gamma$-ray pulsations because typically only one photon is observed per $\sim 10^8$ rotations. We report the first $\gamma$-ray pulsar discovered with an extension of a recently proposed algorithm, which uses efficient tools for identifying short vectors in high-dimensional lattices instead of exhaustively searching for periodic signals. PSR J1622−7204, an isolated pulsar spinning at 261 Hz, was found in 15 years of Fermi-LAT data using about ten CPU hours. In contrast, the previous state-of-the-art (a semicoherent search) failed after $\sim$600 CPU years. This new approach may enable the discovery of many new $\gamma$-ray pulsars, including in binary systems, a previously intractable problem without detailed prior knowledge of the orbital parameters.

*Introduction*—The Fermi Large Area Telescope (LAT) [1] records $\gamma$-ray photon arrival times with microsecond precision. Pulsations have been detected from more than 300 pulsars—rapidly rotating, highly magnetized neutron stars [2]—including the fastest Galactic pulsar [3], the most compact Galactic pulsar binary [4], and the heaviest known neutron stars [5], which constrain the nuclear equation of state [6]. Millisecond pulsars can also form a $\gamma$-ray pulsar timing array that probes the gravitational-wave background without radio-specific systematic uncertainties [7].

Yet, among the thousands of sources in the fourth Fermi-LAT source catalog (4FGL) [8], hundreds with pulsar-like spectra and properties remain unidentified after both radio and $\gamma$-ray pulsation searches. Searching these for $\gamma$-ray pulsations is computationally challenging. With an effective collecting area of $\sim 1\ m^2$, the LAT detects fewer than $\sim 10^3$ photons from a faint pulsar over its 18-year mission, during which the pulsar rotates $10^9$ to $10^{11}$ times. If the detector had $10^8$ times the collecting area, or the source were $10^4$ times closer to us, then most rotations would yield at least one photon, making the rotation period easy to measure. Instead, we observe only a tiny, random subset of photons, contaminated by background. This makes the problem very difficult.

 *Open access publication funded by the Max Planck Society.*

There are additional complications. As the pulsar radiates energy and angular momentum, its rotation slows. Furthermore, the pulse arrival times are modified by the LAT's motion around the Solar System. Accounting for these effects in a brute-force search requires an intractable number ($\sim 10^{22}$) of trials [9].

Existing methods [9,10] trade sensitivity for a $\sim$5-order-of-magnitude reduction in search points, yet still require substantial computing resources, such as those of the Einstein@Home volunteer distributed computing project. Nevertheless, this approach has enabled the discovery of more than eighty pulsars [2,11–16].

Recently, Gazith, Pearlman, and Zackay proposed a new method [17], which in practice reduces the computational cost of a search by a further 5 to 6 orders of magnitude. This is achieved by transforming it into a classic geometry problem: finding the *shortest* nonzero vector in a lattice. A lattice is a regularly spaced grid of points: all integer linear combinations of a set of linearly independent vectors.

Here, we report an extension of this algorithm and its first success. We hunted for pulsations from the unassociated source 4FGL J1622.2−7202 [8], whose spectral and other properties have long suggested it may host a pulsar. We found a new isolated millisecond pulsar J1622−7204 in $\approx 1\%$ of our trial runs, each taking under 5 min on one CPU core. This is by far the least computationally expensive successful blind $\gamma$-ray pulsar search to date. Previous Einstein@Home searches (using the setup of [16]) and TRAPUM radio searches [18,19] both failed to detect pulsations from PSR J1622−7204.

*Method*—We use an improved version of the lattice-based pulsar search presented in [17]. First, we describe how the pulsar search is turned into a lattice problem and discuss our modifications.

*Input data*—The inputs for the lattice algorithm are the photon arrival times (PATs) at the satellite of $n$ photons detected by the Fermi-LAT (with $n = 80$ in our search) from a small sky region around a target source, in this case 4FGL J1622.2−7202. The ordered PATs are denoted by $t_1 < t_2 < \ldots < t_n$. We also need the photon weights $w_i \in [0, 1]$ for $i = 1, \ldots, n$, which are the (prior) probability that the $i$th photon was emitted by the source, rather than the background [20]. The $w_i$ are derived from the photon energies and directions, using a spectral-spatial model of the time-averaged $\gamma$-ray sky.

The PATs are affected by factors both related and unrelated to the pulsar's motion. The largest arise from the LAT's motion around Earth, and the Earth's motion around the Sun. Over a year, these perturb the satellite-to-pulsar distance, creating annual PAT variations of up to ≈500 s (Sun-to-Earth light travel time). To account for these, we compute the times $\tau$ at which photons would arrive at the Solar System barycenter (SSB) if emitted from the pulsar simultaneously with the observed ones. [Barycentric dynamical time (TDB) $\tau$ is defined in the International Celestial Reference System (ICRS) frame, a standard inertial coordinate system with origin at the SSB, and runs at the same average rate as Global Positioning System time at the Earth geoid].

*Pulsar rotation phase model*—The relationship between the PATs $t$ and $\tau$ depends upon the (unknown) direction to the pulsar, given by its right ascension $\alpha$ and declination $\delta$ in the ICRS frame:

$$\alpha = \alpha_0 + \Delta\alpha, \qquad \delta = \delta_0 + \Delta\delta. \tag{1}$$

The nominal direction $\alpha_0$, $\delta_0$ in Table I is a source localization estimate using 14 years of 4FGL data [21]. The arrival times at the SSB are

$$\tau(t,\alpha,\delta) = \tau(t,\alpha_0,\delta_0) + \Delta\alpha\frac{\partial\tau}{\partial\alpha}(t,\alpha_0,\delta_0) + \Delta\delta\frac{\partial\tau}{\partial\delta}(t,\alpha_0,\delta_0), \tag{2}$$

where we linearize in the sky offsets $\Delta\alpha$, $\Delta\delta$. For each PAT, we compute $\tau(t_i, \alpha_0, \delta_0)$ and its partial derivatives using PINT [22–24], which uses the location of the LAT within the Solar System [25]. We set $\tau = 0$ at the reference epoch, modified Julian date (MJD) 57662.23910470847.

We assume that the pulsar is isolated, moving at constant velocity with respect to the SSB, and slowly radiating rotational energy and angular momentum. Its rotation phase is described by a quadratic spin-down

$$\Phi(\tau) \equiv \phi_0 + f\tau + \frac{1}{2}\dot{f}\tau^2, \tag{3}$$

increasing by one for every full rotation. At time zero, $\phi_0 \in [-0.5, 0.5)$ is the initial phase, $f$ is the pulsar spin frequency, and $\dot{f}$ (typically $< 0$) is the spin-down rate. Last, we assume that photons are emitted in pulses centered about the integer values of $\Phi$.

*Pulsar search problem*—Our model implies that PATs lie close to times $\tau$ for which $\Phi(\tau)$ is integer. From those PATs, we want to find the 5-vector

$$\boldsymbol{p} = \left[\phi_0, f, \dot{f}, f\Delta\alpha, f\Delta\delta\right] \tag{4}$$

containing the correct values of the three phase-model parameters and the two (scaled) sky direction offsets. As explained above, this is difficult because we do not know the number of rotations $k_i \in \mathbb{Z}$ associated with the $i$th photon. With some caveats described later, the search problem is solved by minimizing the (squared) residual phases $\varphi_i \equiv \Phi[\tau(t_i, \alpha, \delta)] - k_i$. If $k_i$ is the closest integer to $\Phi[\tau(t_i, \alpha, \delta)]$, then $\varphi_i \in [-0.5, 0.5]$. Once the integer vector $\boldsymbol{k} \equiv [k_1, \ldots, k_n]$ is known, it is easy to find $\boldsymbol{p}$. Note that in practice the PATs are widely separated, so physical solutions must have increasing $\boldsymbol{k}$, meaning increasing rotation counts $k_1 < k_2 < \ldots < k_n$.

We express the search problem in linear form

$$\boldsymbol{v} \equiv \boldsymbol{u}\boldsymbol{B}, \tag{5}$$

where the full-rank $(n+5) \times (n+5)$ matrix

$$\boldsymbol{B} \equiv \begin{bmatrix} -\boldsymbol{I}_n & \boldsymbol{0} \\ \boldsymbol{S} & \boldsymbol{\eta} \end{bmatrix} \tag{6}$$

is computed at the start of the search, and the row vector $\boldsymbol{u} = [\boldsymbol{k}\ \boldsymbol{p}]$ contains the $n+5$ unknowns. (Notation: $[\boldsymbol{x}\,\boldsymbol{y}]$ concatenates the $n$ columns of $\boldsymbol{x}$ before the 5 columns of $\boldsymbol{y}$.) Here, $\boldsymbol{I}_n$ is the $n \times n$ identity matrix, $\boldsymbol{S}$ is the $5 \times n$ matrix

$$\boldsymbol{S} \equiv \begin{bmatrix} 1 & \cdots & 1 \\ \tau(t_1,\alpha_0,\delta_0) & \cdots & \tau(t_n,\alpha_0,\delta_0) \\ \frac{1}{2}\tau^2(t_1,\alpha_0,\delta_0) & \cdots & \frac{1}{2}\tau^2(t_n,\alpha_0,\delta_0) \\ \frac{\partial\tau}{\partial\alpha}(t_1,\alpha_0,\delta_0) & \cdots & \frac{\partial\tau}{\partial\alpha}(t_n,\alpha_0,\delta_0) \\ \frac{\partial\tau}{\partial\delta}(t_1,\alpha_0,\delta_0) & \cdots & \frac{\partial\tau}{\partial\delta}(t_n,\alpha_0,\delta_0) \end{bmatrix}, \tag{7}$$

$\boldsymbol{0}$ is a zero matrix of appropriate shape [in (6), $n \times 5$], and $\boldsymbol{\eta}$ is the $5 \times 5$ diagonal matrix

$$\boldsymbol{\eta} \equiv \mathrm{diag}\left(\frac{\sigma}{\sigma_{\phi_0}}, \frac{\sigma}{\sigma_f}, \frac{\sigma}{\sigma_{\dot{f}}}, \frac{\sigma}{\sigma_{f\Delta\alpha}}, \frac{\sigma}{\sigma_{f\Delta\delta}}\right). \tag{8}$$

The subscripted $\sigma$'s in Table I are the widths of zero-mean Gaussian priors on the search parameters $\boldsymbol{p}$ in (4); $\sigma = 0.1$ rotations is the assumed Gaussian pulse width. The priors $\sigma_{f\Delta\alpha}, \sigma_{f\Delta\delta}$ define a sky patch slightly larger than the 4FGL 95% confidence radius for $\alpha_0$, $\delta_0$ at 300 Hz.

TABLE I. Search parameters.

| | | |
|---|---|---|
| PATs used for lattice | $n$ | 80 |
| Nominal right ascension (RA) | $\alpha_0$ | $16^h22^m12\overset{s}{.}72$ |
| Nominal declination (DEC) | $\delta_0$ | $-72°03'57\overset{''}{.}6$ |
| Assumed pulse width (duty cycle) | $\sigma$ | 0.1 |
| Initial phase prior width | $\sigma_{\phi_0}$ | 1 |
| Spin frequency prior width | $\sigma_f$ | 300 Hz |
| Spin-down prior width | $\sigma_{\dot{f}}$ | $10^{-14}$ Hz s$^{-1}$ |
| Scaled RA prior width | $\sigma_{f\Delta\alpha}$ | $(0^h0^m12\overset{s}{.}0)\sigma_f$ |
| Scaled DEC prior width | $\sigma_{f\Delta\delta}$ | $(0°03'00\overset{''}{.}0)\sigma_f$ |

We now show that via (5), the row vector $\boldsymbol{u}$ containing the correct pulsar parameters and rotation counts produces a row vector $\boldsymbol{v}$ with relatively short length $\|\boldsymbol{v}\|$. This transforms the pulsar search problem into a form similar to the classic "shortest vector problem" (SVP) in lattice theory.

Consider the first entry $v_1$ of (5), dropping higher orders in $\Delta\alpha, \Delta\delta$, and assuming $|\dot{f}|$ is small enough that $|\tau\dot{f}(\Delta\alpha\partial\tau/\partial\alpha + \Delta\delta\partial\tau/\partial\delta)| \sim |\dot{f}|(5500\ \mathrm{d})(500\ \mathrm{s})(1\times\mathrm{rad}^{-1})(3\ \mathrm{arcmin}) \sim |\dot{f}|10^8\mathrm{s/Hz} \ll 1$. Inspection of (6) and (7) gives $v_1 = -k_1 + \Phi[\tau(t_1,\alpha,\delta)]$. For the correct $\boldsymbol{u} = [\boldsymbol{k}\ \boldsymbol{p}]$, this gives $v_1/\sigma = \varphi_1/\sigma$: the normalized residual phase of the first photon, of order unity. This also holds for the next $n-1$ components (photons).

Next, consider the $(n+1)$th entry of (5). Inspection of (6) and (8) gives $v_{n+1}/\sigma = \phi_0/\sigma_{\phi_0}$. This "penalty" for the first search parameter $\phi_0$ is of order unity if $|\phi_0| \lesssim \sigma_{\phi_0}$, i.e., if $\phi_0$ lies within the "prior range" of Table I. This also holds for the final four components (search parameters).

Thus, dividing $\|\boldsymbol{v}\|^2$ by $\sigma^2$,

$$\frac{\|\boldsymbol{v}\|^2}{\sigma^2} = \frac{\phi_0^2}{\sigma_{\phi_0}^2} + \frac{f^2}{\sigma_f^2} + \frac{\dot{f}^2}{\sigma_{\dot{f}}^2} + \frac{f^2\Delta\alpha^2}{\sigma_{f\Delta\alpha}^2} + \frac{f^2\Delta\delta^2}{\sigma_{f\Delta\delta}^2} + \sum_{i=1}^{n}\frac{\varphi_i^2}{\sigma^2}. \quad (9)$$

This will be $\approx n+5$ for the correct integer rotation counts $\boldsymbol{k}$ and pulsar parameters $\boldsymbol{p}$, and much larger for the wrong values, provided that the pulse profile has width $\approx \sigma$. Thus, the solution to the search problem corresponds to parameters $\boldsymbol{u}$ that lead to "short" vectors $\boldsymbol{v}$.

If the phase residuals and model parameters are assumed to have been drawn from independent zero-mean Gaussians with the widths in Table I, then $-\|\boldsymbol{v}\|^2/2\sigma^2$ is the logarithm of the posterior probability [17]. Short vectors therefore correspond to solutions with high posterior probability.

The algorithm presented in an earlier work [17] discretizes the five continuous real parameters $\boldsymbol{p}$, transforming the search into an $(n+5)$-dimensional SVP. We improve this with an additional step.

*Lattice formulation*—Below, we solve for the five real parameters $\boldsymbol{p}$ of $\boldsymbol{u} = [\boldsymbol{k}\ \boldsymbol{p}]$, reducing the dimension from $n+5$ to $n$. The remaining unknowns $\boldsymbol{k}$ are then found by solving the SVP. From (5) and (6), for fixed $\boldsymbol{k}$ one has

$$\boldsymbol{v} = \boldsymbol{u}\boldsymbol{B} = [-\boldsymbol{k}\ \ \boldsymbol{0}] + \boldsymbol{p}\boldsymbol{W}, \quad (10)$$

where $\boldsymbol{W} \equiv [\boldsymbol{S}\ \boldsymbol{\eta}]$ is the last five rows of $\boldsymbol{B}$. As $\boldsymbol{p}$ varies, $\boldsymbol{v}$ traces out a 5-plane in $\mathbb{R}^{n+5}$. The shortest $\boldsymbol{v}$ is obtained when $\boldsymbol{v}$ is orthogonal to that 5-plane. Thus, the vector $\boldsymbol{p} = \boldsymbol{p}_{\min}$ that minimizes $\|\boldsymbol{v}\|$ satisfies the five equations

$$\boldsymbol{v}\boldsymbol{W}^T = ([-\boldsymbol{k}\ \ \boldsymbol{0}] + \boldsymbol{p}_{\min}\boldsymbol{W})\boldsymbol{W}^T = \boldsymbol{0}, \quad (11)$$

since the dot product between $\boldsymbol{v}$ in (10) and any vectors in the 5-plane (here, the five columns of the transpose $\boldsymbol{W}^T$) must vanish. Multiplying (11) on the right by $(\boldsymbol{W}\boldsymbol{W}^T)^{-1}$ and moving the $[-\boldsymbol{k}\ \boldsymbol{0}]$ term to the rhs gives

$$\boldsymbol{p}_{\min} = [\boldsymbol{k}\ \ \boldsymbol{0}]\boldsymbol{W}^T(\boldsymbol{W}\boldsymbol{W}^T)^{-1}. \quad (12)$$

Define the $(n+5)\times(n+5)$ symmetric matrix

$$\boldsymbol{P} \equiv -\boldsymbol{I}_{n+5} + \boldsymbol{W}^T(\boldsymbol{W}\boldsymbol{W}^T)^{-1}\boldsymbol{W}, \quad (13)$$

noting that $-\boldsymbol{P}$ is a projection operator, since $\boldsymbol{P}^2 = -\boldsymbol{P}$. Substituting $\boldsymbol{p}_{\min}$ from (12) into (10) gives

$$\boldsymbol{v} = [\boldsymbol{k}\ \boldsymbol{0}]\boldsymbol{P} = \boldsymbol{k}\boldsymbol{P}^{\perp}, \quad (14)$$

where $\boldsymbol{P}^{\perp}$ is obtained from $\boldsymbol{P}$ by removing the last five rows.

The set of vectors $\boldsymbol{v} = \boldsymbol{k}\boldsymbol{P}^{\perp}$ with integer $\boldsymbol{k} \in \mathbb{Z}^n$ forms a lattice with generator matrix $\boldsymbol{P}^{\perp}$. Finding a nonzero integer $\boldsymbol{k}$ that minimizes the length of $\boldsymbol{v}$ is called the SVP. The matrix $\boldsymbol{P}^{\perp}$ is rectangular: $n\times(n+5)$. Equivalent square generator matrices may be found from (14), which implies $\|\boldsymbol{v}\|^2 = \boldsymbol{k}\boldsymbol{Q}\boldsymbol{k}^T$. The symmetric positive-definite $n\times n$ Gram matrix $\boldsymbol{Q} \equiv \boldsymbol{P}^{\perp}(\boldsymbol{P}^{\perp})^T$ has a unique positive definite symmetric square-root $\boldsymbol{M}$ satisfying $\boldsymbol{M}^2 = \boldsymbol{Q}$. From $\|\boldsymbol{v}\|^2 = \|\boldsymbol{k}\boldsymbol{M}\|^2$, it follows that $\boldsymbol{M}$ is an $n\times n$ generator matrix for the (possibly rotated) lattice generated by $\boldsymbol{P}^{\perp}$. Alternatively, Cholesky decomposition produces a square lower triangular lattice generator matrix $\boldsymbol{L}$, with $\boldsymbol{Q} = \boldsymbol{L}\boldsymbol{L}^T$ implying that $\|\boldsymbol{v}\|^2 = \|\boldsymbol{k}\boldsymbol{L}\|^2$.

With this formulation, we do not search over the five parameters $\boldsymbol{p}$. Instead, we search for rotation count vectors $\boldsymbol{k}$ that yield short vectors in the equivalent lattices generated by $\boldsymbol{P}^{\perp}$, $\boldsymbol{M}$, or $\boldsymbol{L}$. The corresponding $\boldsymbol{p}$ follow from (12).

Our approach differs from [17] in that we analytically extremize $\boldsymbol{p}$ and eliminate it, thereby reducing the dimension. In contrast, [17] works with the lattice generated by $\boldsymbol{B}$, after rescaling $\boldsymbol{p}$ so that it takes discrete integer values. This requires an inverse rescaling of the rows of $\boldsymbol{W}$ in order to keep the product $\boldsymbol{p}\boldsymbol{W}$ in (10) invariant. The rescaled rows of $\boldsymbol{W}$ have very small Euclidean norms, leading to a large number of short but unphysical lattice vectors with small integers in (the rescaled) $\boldsymbol{p}$ and $\boldsymbol{k} = \boldsymbol{0}$. This issue is resolved using projections [different than our $-\boldsymbol{P}$ defined by (13)]. In [17], Sec. 4 and the Appendix, the blocks $\boldsymbol{S}$ and $\boldsymbol{\eta}$ of $\boldsymbol{W}$ are orthogonalized separately, thereby changing the subspace spanned by the last five rows of $\boldsymbol{B}$, which is then projected out. We are still studying the advantages and disadvantages of the lattice that [17] would produce, which differs from the one presented here.

*Short vectors versus shortest vectors*—In practice, the shortest vectors $\boldsymbol{v}$ in the lattice come [via (14)] from rotation counts $\boldsymbol{k}$ that are incorrect "solutions" to the search problem. Many of these $\boldsymbol{k}$ have parameters $\boldsymbol{p}$ [via (12)] that are unphysical. Examples are all $k_i$ being zero, except for one or a few being $\pm 1$, or nonmonotonic $\boldsymbol{k}$, or $|f| \lesssim 1$ mHz. The large trials factor also means that spurious short vectors may arise by chance. Furthermore, our data contain spurious background PATs, and most realistic pulsars have pulse profiles that are more complicated than our assumed Gaussian shape, lengthening $\boldsymbol{v}$ for the correct $\boldsymbol{p}$ (e.g., the pulse profile of the pulsar we discovered has two peaks). Similar nonsolutions appear in [17] and we address this the same way, i.e., by generating many short vectors and identifying the correct solution among them. We are currently optimizing the method for more complicated pulse profiles, but defer this discussion to future work.

The correct solution $\boldsymbol{p}$ is identified by testing the different possible solutions using the $H$ statistic [20,26], normalized as in [13], Eqs. (7)–(11). The $H$ statistic is computed for the highest-weighted 1000 PATs, using the residual phases determined by $\boldsymbol{p}$. If $H > 150$, then the parameters $\boldsymbol{p}$ are recorded as a potential solution. The corresponding false alarm probability (for many independent random PATs) is $p \approx \exp(-0.4\ H) \approx 10^{-26}$, whose inverse is larger than the number of independent points in the parameter space ($\sim 10^{22}$), making the trials-corrected significance similar to the standard $4.1\sigma$ detection threshold [27].

*Implementation*—Several algorithms and libraries have been developed to solve the classic SVP, but we require *many* short vectors. We use a recent record holder in the SVP challenge, the G6K library [28]. This employs sieving, which repeatedly forms sums and differences of lattice vectors to produce shorter ones.

G6K requires integer generators, so we multiply $\boldsymbol{L}$ by $10^{30}$ and round to the nearest integers, following [17]. While one could sieve this basis directly, it is more efficient to first apply block Korkine-Zolotarev reduction and then invoke G6K's "pump-up" procedure [28]. This begins by projecting the lattice orthogonally to the first $n/2$ basis vectors and sieving the resulting $n/2$-dimensional lattice. The projection dimension is then increased by adding one additional basis vector, after which projection and sieving are repeated. This continues until the full $n$-dimensional lattice is reached. Throughout, G6K maintains and enlarges a database of short vectors. At the end, these are transformed back to the original basis, yielding many $\boldsymbol{k}$ with short $\boldsymbol{kL}$.

TABLE II. Vectors obtained from sieving with G6K.

| | |
|---|---|
| Vectors in database | 318 199 |
| Unphysical vectors (nonincreasing $\boldsymbol{k}$) | 1434 |
| Equivalent vectors removed | 126 244 |
| Solutions ($H$ statistic $H > 150$) | 1 ($H \approx 231$) |
| Shortest vector norm | $\lVert\boldsymbol{v}\rVert = 0.099\,98$ |
| Shortest physical vector norm | $\lVert\boldsymbol{v}\rVert = 1.093\,36$ |
| Longest vector norm | $\lVert\boldsymbol{v}\rVert = 1.311\,85$ |
| Solution vector norm | $\lVert\boldsymbol{v}\rVert = 1.212\,15$ |
| Database vectors shorter than solution | 6827 |

We postprocess the vectors by flipping the signs of any $\boldsymbol{k}$ with decreasing values, removing any unphysical vectors, and also removing any equivalent entries, where $\boldsymbol{k}$ and $\boldsymbol{k}'$ differ by the same integer. Then, we examine the $H$ statistic as described above. A notebook reproducing our method is available in [29].

*Dataset*—We selected photons detected before MJD 60284 from the unassociated source 4FGL J1622.2−7202 [8]. We used Pass 8 P8R3 Fermi-LAT data [30,31] and the `P8R3_SOURCE_V2` instrument response functions, selecting `SOURCE`-class photons with energies above 100 MeV, and used the energy-dependent angular and point spread function event-type cuts described in [8]. We computed probability weights $w_i$ with `gtsrcprob` [32], using the 4FGL-DR4 model [21] for the spectra and positions of the target and nearby sources, and the corresponding interstellar (`gll_iem_v07.fits`) and isotropic (`iso_P8R3_SOURCE_V2_v1.txt`) diffuse emission models [33].

*Multiple trials and random photon subsets*—The computational cost of the sieving process grows exponentially with $n$. While including more photons should increase sensitivity, the algorithm is very sensitive to outlier PATs with large residual phases $|\varphi_i|$ (background photons). To mitigate these effects, we start with the highest-weighted 120 photons, pick random subsets of size $n = 80$ from these, and then carry out the lattice-based analysis described above. This is an extension of the method suggested in [17]. A search using a single random subset of PATs takes about 90 s to run on a single CPU core of an otherwise idle AMD EPYC 7351P 16-Core processor (this extends to $\sim 5$ min under full load).

*First discovery*—The pulsar PSR J1622−7204 was found the first time that we tried our algorithm on 20 different random subsets. The search parameters are given in Table I. By chance, this trial used PATs containing only a few outliers. For other subsets containing a larger number of outliers, the algorithm failed to find the pulsar.

We got lucky with the 20 trials, of which one was successful. Based on 1000 trials, we estimate that the probability of successfully detecting this pulsar in a single subset of 80 PATs is $p \approx 1\%$. With $N$ trials, the probability of finding the pulsar at least once is $1 - (1 - p)^N$, which is 18% for 20 trials, and 50% for 70 trials. A robust detection with 95% success probability is achieved for 300 trials.

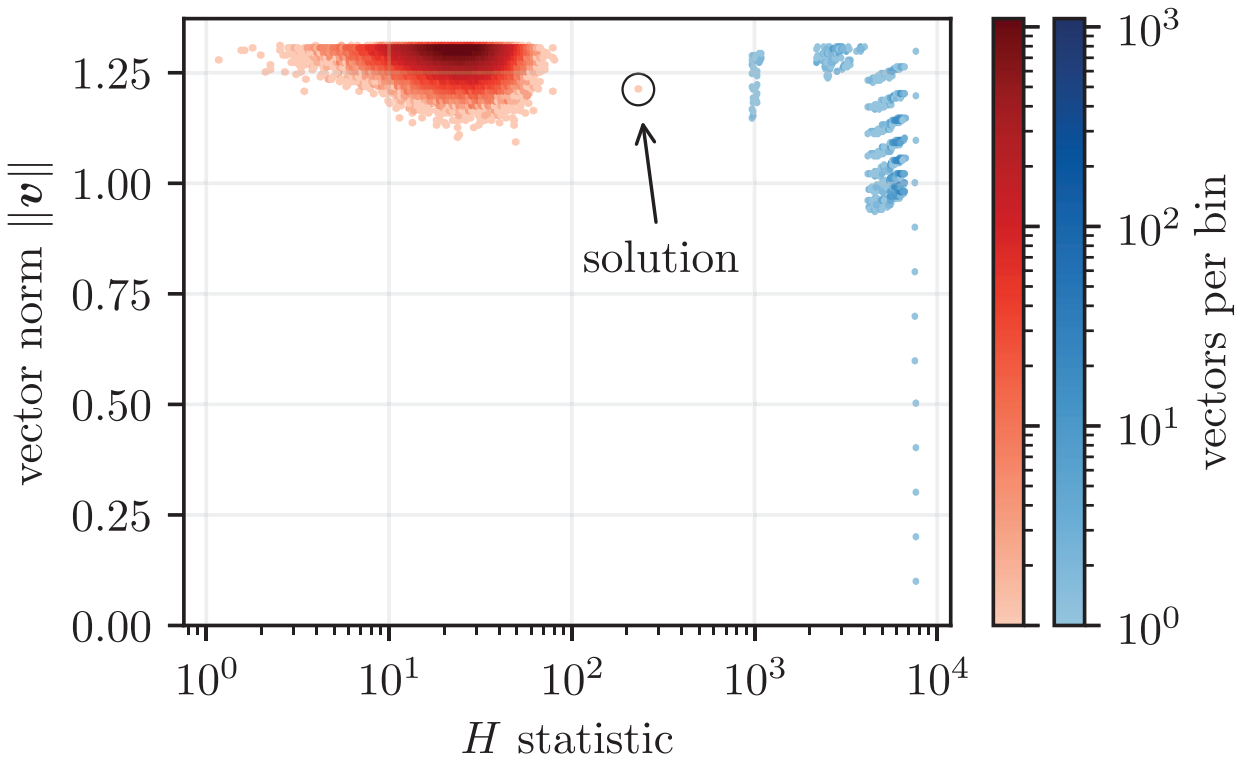

FIG. 1. Density of the 318 199 short vectors found by G6K in the successful search, i.e., for one subset of 80 photons. The (correct) solution is circled. Vectors (blue) with $H$ statistic larger than the solution are unphysical: they have frequencies $f < 10^{-7}$ Hz and their $\boldsymbol{k}$ are not increasing. The remaining vectors (red) have increasing $\boldsymbol{k}$ and frequencies 3 mHz $< f <$ 2 kHz.

Table II contains information about the shortest G6K vectors of the successful trial. The set of short vectors is illustrated in Fig. 1 in the plane of vector norm versus $H$ statistic. The two groups of nonsolution vectors, the "physical" and "unphysical" ones, appear in all of our searches, independent of the source properties or its pulse profile. Even after the unphysical vectors are removed, the solution is not the shortest vector. However, its $H$ statistic $H = 231$ has a $p$-value $p \approx \exp(-0.4\,H) \approx 10^{-40}$ [20]. This is highly significant, since $1/p$ is much larger than the number of independent trial points in parameter space, $\sim 10^{22}$.

*Pulsar parameters*—The physical properties of PSR J1622−7204 are summarized in Table III. These parameters and their uncertainties were determined from the 15 905 photons with weights $w_i > 0.01$, using the $\gamma$-ray pulsar timing method described in [34] to explore the likelihood surface in the neighborhood of the physical lattice vector with the highest $H$ statistic. These parameters and derived physical properties are typical for an isolated $\gamma$-ray millisecond pulsar (see, e.g., Fig. 2 of Ref. [2]). The uncertainties are relatively small because of the long data span of ~15 years. Figure 2 shows the pulse profile and the 15 905 weighted photon phases over time. The effective number of photons from the pulsar is $\sum_i w_i \approx 828$, and the expected squared pulsation signal-to-noise ratio is $\sum_i w_i^2 \approx 193$.

*Conclusions*—We have shown that lattice-based pulsar searches can discover previously undetected pulsars using 5 to 6 orders of magnitude less computing time than prior methods. The new method will be an important tool in future $\gamma$-ray pulsar searches.

PSR J1622−7204 is an isolated millisecond pulsar. An exciting prospect is that the lattice-based algorithm can also be used for blind searches for pulsars in binary systems. These are computationally infeasible with previous methods [9]. The Fermi-LAT source catalog [8] contains hundreds of unassociated sources whose spectral shape and temporal variability suggest that they might be pulsars [35], but searches for isolated pulsars have failed. Many of these could be $\gamma$-ray pulsars in binary systems. Searching for binary pulsars requires modeling the additional PAT delays due to the pulsar's orbital motion, which produces additional rows in (7) and (8). A blind search for a pulsar in a binary system should have a computation cost $\sim 10^6$ larger than the isolated search reported here [17], since it requires linearization around $\sim 10^6$ binary parameter sets, each defining a different lattice. This should be feasible with Einstein@Home or similar resources. To compensate for the increased trials factor and search volume, we estimate that the threshold on $H$ increases by ~71. We do not expect this to be an issue in practice, since many of the potential sources have enough high-weight photons to result in a much larger $H$ statistic.

TABLE III. Properties of PSR J1622−7204.

| | |
|---|---|
| **Units and conventions** | |
| Time coordinate $\tau$ | TDB |
| Reference epoch $\tau = 0$ | MJD 57662.23910470847 |
| Solar System ephemeris model | DE440 |
| Data span | MJD 54 694 to 60 284 |
| **Inferred parameters of PSR J1622−7204** | |
| Spin frequency $f$ | 260.896 208 070 72(2) Hz |
| Frequency derivative $\dot{f}$ | $-1.5845(4) \times 10^{-15}$ Hz/s |
| Right ascension (RA) $\alpha$ | $16^h22^m16\overset{s}{.}553(2)$ |
| Declination (DEC) $\delta$ | $-72°04'25\overset{''}{.}27(1)$ |
| Proper motion in RA, $\mu_\alpha \cos\delta$ | −2.9(1.6) mas/yr |
| Proper motion in DEC, $\mu_\delta$ | −4.7(2.1) mas/yr |
| **Derived parameters of PSR J1622−7204** | |
| Spin period | 3.832 941 871 385 6(3) ms |
| Spin-period derivative | $2.3279(6) \times 10^{-20}$ |
| Characteristic age[a] | $2.6 \times 10^9$ yr |
| Surface magnetic field[a] | $3.0 \times 10^8$ G |
| Light-cylinder magnetic field[a] | $4.9 \times 10^4$ G |
| Spin-down power[a] | $1.6 \times 10^{34}$ erg/s |

[a]These parameters are derived from the observed spin-period derivative, which includes contributions from the Shklovskii effect and Galactic acceleration relative to the SSB. We cannot correct for these (unknown) distance-dependent effects, but both are small and approximately cancel, leaving a net correction smaller than ±6%.

*Acknowledgments*—We thank Toby Burnett for identifying 4FGL J1622.2−7202 as a promising pulsar candidate source for our searches, and M. A. Papa for helpful comments on the manuscript. B. Z. and D. G. are supported by the Minerva Foundation with funding from the Federal

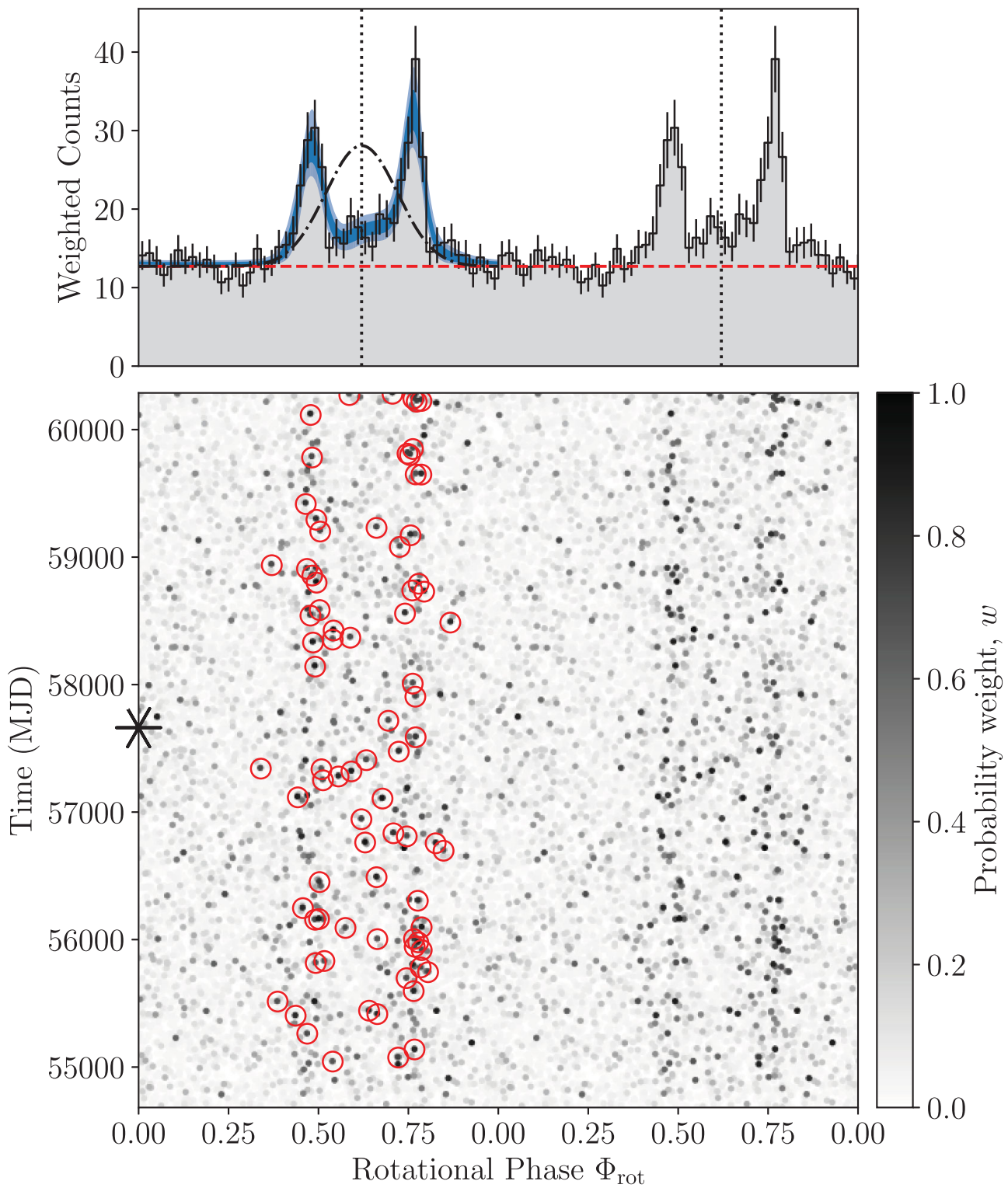

FIG. 2. Bottom: weighted photon phases for the best-fit timing solution, showing two rotations. The left-axis star marks $\tau = 0$, the reference epoch of Table III. The rotational phase $\Phi_{\rm rot}(\tau) \equiv [\Phi(\tau) - \phi_0] \bmod 1$ with $\phi_0 \approx 0.38$ is offset from (3) so that $\Phi_{\rm rot}(0) = 0$. The 80 successful trial PATs are circled in red. Top: integrated pulse profile, with dark and light blue showing $1\sigma/2\sigma$ posteriors. The dashed red line estimates the nonpulsed background. The assumed $\sigma = 0.1$ Gaussian pulse profile (dash-dot) is centered around integer $\Phi(\tau)$, marked by vertical lines, where the lattice algorithm identified peaks.

German Ministry for Education and Research. The Fermi-LAT Collaboration acknowledges support for LAT development, operation, and data analysis from NASA and the U.S. Department of Energy; CEA/Irfu and IN2P3/CNRS in France; ASI and INFN in Italy; MEXT, KEK, and JAXA in Japan; and the K. A. Wallenberg Foundation, the Swedish Research Council, and the Swedish National Space Board. Science analysis support in the operations phase from INAF (Italy) and CNES (France) is also gratefully acknowledged. That work was performed in part under DOE Contract No. DE-AC02-76SF00515.

*Data availability*—The code to reproduce our result is available in [29] and includes a notebook implementing the lattice-based search as well as a notebook performing the parameter estimation. The resulting parameter file `data/J1622-7204_shoogle_results.par` (also included in [29]) can be loaded into PINT together with the PATs to produce a pulse profile compatible with Fig. 2 as well as an $H$ statistic of ~345.

[1] W. B. Atwood, A. A. Abdo, M. Ackermann, W. Althouse, B. Anderson, M. Axelsson, L. Baldini, J. Ballet, D. L. Band, G. Barbiellini *et al.*, The Large Area Telescope on the Fermi Gamma-ray Space Telescope mission, Astrophys. J. **697**, 1071 (2009).

[2] D. A. Smith, S. Abdollahi, M. Ajello, M. Bailes, L. Baldini, J. Ballet, M. G. Baring, C. Bassa, J. Becerra Gonzalez, R. Bellazzini *et al.*, The third Fermi Large Area Telescope catalog of gamma-ray pulsars, Astrophys. J. **958**, 191 (2023).

[3] C. G. Bassa, Z. Pleunis, J. W. T. Hessels, E. C. Ferrara, R. P. Breton, N. V. Gusinskaia, V. I. Kondratiev, S. Sanidas, L. Nieder, C. J. Clark *et al.*, LOFAR discovery of the fastest-spinning millisecond pulsar in the Galactic field, Astrophys. J. Lett. **846**, L20 (2017).

[4] L. Nieder, C. J. Clark, D. Kandel, R. W. Romani, C. G. Bassa, B. Allen, A. Ashok, I. Cognard, H. Fehrmann, P. Freire *et al.*, Discovery of a gamma-ray black widow pulsar by GPU-accelerated Einstein@Home, Astrophys. J. Lett. **902**, L46 (2020).

[5] R. W. Romani, M. Beleznay, A. V. Filippenko, T. G. Brink, and W. Zheng, PSR J0952-0607: Tightening a record-high neutron star mass, Astrophys. J. **996**, 101 (2026).

[6] F. Özel and P. Freire, Masses, radii, and the equation of state of neutron stars, Annu. Rev. Astron. Astrophys. **54**, 401 (2016).

[7] M. Ajello, W. B. Atwood, L. Baldini, J. Ballet, G. Barbiellini, D. Bastieri, R. Bellazzini, A. Berretta, B. Bhattacharyya *et al.* (FERMI-LAT Collaboration), A gamma-ray pulsar timing array constrains the nanohertz gravitational wave background, Science **376**, 521 (2022).

[8] S. Abdollahi, F. Acero, M. Ackermann, M. Ajello, W. B. Atwood, M. Axelsson, L. Baldini, J. Ballet, G. Barbiellini, D. Bastieri *et al.*, Fermi Large Area Telescope fourth source catalog, Astrophys. J. Suppl. Ser. **247**, 33 (2020).

[9] L. Nieder, B. Allen, C. J. Clark, and H. J. Pletsch, Exploiting orbital constraints from optical data to detect binary gamma-ray pulsars, Astrophys. J. **901**, 156 (2020).

[10] W. B. Atwood, M. Ziegler, R. P. Johnson, and B. M. Baughman, A time-differencing technique for detecting radio-quiet gamma-ray pulsars, Astrophys. J. **652**, L49 (2006).

[11] A. A. Abdo, M. Ackermann, M. Ajello, B. Anderson, W. B. Atwood, M. Axelsson, L. Baldini, J. Ballet, G. Barbiellini, M. G. Baring *et al.*, Detection of 16 gamma-ray pulsars through blind frequency searches using the Fermi LAT, Science **325**, 840 (2009).

[12] P. M. Saz Parkinson, M. Dormody, M. Ziegler, P. S. Ray, A. A. Abdo, J. Ballet, M. G. Baring, A. Belfiore, T. H. Burnett, G. A. Caliandro *et al.*, Eight $\gamma$-ray pulsars discovered in blind frequency searches of Fermi LAT data, Astrophys. J. **725**, 571 (2010).

[13] H. J. Pletsch, L. Guillemot, B. Allen, M. Kramer, C. Aulbert, H. Fehrmann, P. S. Ray, E. D. Barr, A. Belfiore, F. Camilo *et al.*, Discovery of nine gamma-ray pulsars in Fermi Large Area Telescope data using a new blind search method, Astrophys. J. **744**, 105 (2012).

[14] H. J. Pletsch, L. Guillemot, B. Allen, D. Anderson, C. Aulbert, O. Bock, D. J. Champion, H. B. Eggenstein, H. Fehrmann, D. Hammer *et al.*, Einstein@Home discovery

of four young gamma-ray pulsars in Fermi LAT data, Astrophys. J. Lett. **779**, L11 (2013).

[15] C. J. Clark *et al.*, The Einstein@Home gamma-ray pulsar survey. I. Search methods, sensitivity and discovery of new young gamma-ray pulsars, Astrophys. J. **834**, 106 (2017).

[16] C. J. Clark, M. Di Mauro, J. Wu, B. Allen, O. Behnke, H. B. Eggenstein, B. Machenschalk, L. Nieder, P. M. Saz Parkinson, A. Ashok *et al.*, Einstein@Home searches for gamma-ray pulsars in the inner Galaxy, Astrophys. J. **994**, 149 (2025).

[17] D. Gazith, A. B. Pearlman, and B. Zackay, Recovering pulsar periodicity from time-of-arrival data by finding the shortest vector in a lattice, Astrophys. J. **979**, 48 (2025).

[18] C. J. Clark, R. P. Breton, E. D. Barr, M. Burgay, T. Thongmeearkom, L. Nieder, S. Buchner, B. Stappers, M. Kramer, W. Becker *et al.*, The TRAPUM L-band survey for pulsars in Fermi-LAT gamma-ray sources, Mon. Not. R. Astron. Soc. **519**, 5590 (2023).

[19] T. Thongmeearkom, C. J. Clark, R. P. Breton, M. Burgay, L. Nieder, O. G. Dodge, B. McGloughlin, E. D. Barr, S. Buchner, B. W. Stappers *et al.*, Pulsar discoveries from the TRAPUM UHF survey of Fermi-LAT sources, Mon. Not. R. Astron. Soc. **547**, stag406 (2026).

[20] M. Kerr, Improving sensitivity to weak pulsations with photon probability weighting, Astrophys. J. **732**, 38 (2011).

[21] J. Ballet, P. Bruel, T. H. Burnett, and B. Lott (Fermi-LAT Collaboration), Fermi Large Area Telescope fourth source catalog data release 4 (4FGL-DR4), arXiv:2307.12546.

[22] J. Luo, S. Ransom, P. Demorest, P. S. Ray, A. Archibald, M. Kerr, R. J. Jennings, M. Bachetti, R. van Haasteren, C. A. Champagne *et al.*, PINT: A modern software package for pulsar timing, Astrophys. J. **911**, 45 (2021).

[23] A. Susobhanan, D. L. Kaplan, A. M. Archibald, J. Luo, P. S. Ray, T. T. Pennucci, S. M. Ransom, G. Agazie, W. Fiore, B. Larsen *et al.*, PINT: Maximum-likelihood estimation of pulsar timing noise parameters, Astrophys. J. **971**, 150 (2024).

[24] J. Luo, S. Ransom, P. Demorest, R. van Haasteren, P. Ray, K. Stovall, M. Bachetti, A. Archibald, M. Kerr, J. Colen *et al.*, PINT: High-precision pulsar timing analysis package, Astrophysics Source Code Library, record ascl:1902.007 (2019), https://ascl.net/1902.007.

[25] Fermi-LAT Collaboration, Fermi data & products, https://fermi.gsfc.nasa.gov/ssc/data/ (2026), accessed: 2026-2-8.

[26] O. C. de Jager, J. W. H. Swanepoel, and B. C. Raubenheimer, A powerful test for weak periodic signals with unknown light curve shape in sparse data, Astron. Astrophys. **221**, 180 (1989), https://ui.adsabs.harvard.edu/abs/1989A%26A...221..180D/abstract.

[27] D. A. Smith, P. Bruel, I. Cognard, A. D. Cameron, F. Camilo, S. Dai, L. Guillemot, T. J. Johnson, S. Johnston, M. J. Keith *et al.*, Searching a thousand radio pulsars for gamma-ray emission, Astrophys. J. **871**, 78 (2019).

[28] M. R. Albrecht, L. Ducas, G. Herold, E. Kirshanova, E. W. Postlethwaite, and M. Stevens, The General Sieve Kernel and new records in lattice reduction, in *Advances in Cryptology—EUROCRYPT 2019*, edited by Y. Ishai and V. Rijmen (Springer International Publishing, Cham, 2019), pp. 717–746, 10.1007/978-3-030-17656-3_25.

[29] See Supplemental Material at http://link.aps.org/supplemental/10.1103/vlt2-xzgk for an example implementation of our method and the data and notebook for reproducing our timing solution.

[30] W. Atwood, A. Albert, L. Baldini, M. Tinivella, J. Bregeon, M. Pesce-Rollins, C. Sgrò, P. Bruel, E. Charles, A. Drlica-Wagner *et al.*, Pass 8: Toward the full realization of the Fermi-LAT scientific potential, in *Proceedings of the 4th Fermi Symposium, Monterey, California, 2012*, edited by T. J. Brandt, N. Omodei, and C. Wilson-Hodge, eConf C121028 (2013), pp. 8–13, arXiv:1303.3514.

[31] P. Bruel, T. H. Burnett, S. W. Digel, G. Jóhannesson, N. Omodei, and M. Wood, Fermi-LAT improved Pass 8 event selection, arXiv:1810.11394.

[32] Fermi-LAT Collaboration, Fermitools, v2.2.0, https://github.com/fermi-lat/ScienceTools/releases/tag/Fermitools-2.2.0, https://fermi.gsfc.nasa.gov/ssc/data/analysis/documentation/ (2022).

[33] Fermi-LAT Collaboration, Fermi LAT background models, https://fermi.gsfc.nasa.gov/ssc/data/access/lat/BackgroundModels.html (2024), accessed: 2026-1-12.

[34] C. J. Clark, S. Valtolina, L. Nieder, and R. van Haasteren, Timing gamma-ray pulsars using Gibbs sampling, Astron. Astrophys. **709**, A215 (2026).

[35] D. V. Malyshev, Towards resolving the Galactic center GeV excess with millisecond-pulsars-like sources using machine learning, Phys. Rev. D **111**, 043033 (2025).